\pdfoutput=1
\documentclass[11pt]{article}

\usepackage[final]{acl}

\usepackage{times}
\usepackage{latexsym}

\usepackage[T1]{fontenc}
\usepackage[utf8]{inputenc}

\usepackage{enumitem}
\setlist{nosep}
\usepackage{microtype}

\usepackage{inconsolata}

\usepackage{graphicx}
\usepackage{makecell}
\usepackage{amsmath}
\usepackage{booktabs}

\title{Creating and Evaluating K-12 GenAI Assessment Graders Through Context Engineering}

\author{
  \textbf{Zewei (Victor) Tian\textsuperscript{1}},
  \textbf{Alex Liu\textsuperscript{1}},
  \textbf{Lief Esbenshade\textsuperscript{1}},
  \textbf{Michael Xiao\textsuperscript{1}}
\\
  \textbf{Zachary Zhang\textsuperscript{2}},
  \textbf{Yulia Lápicus\textsuperscript{2}},
  \textbf{Thomas Han\textsuperscript{2}},
  \textbf{Kevin He\textsuperscript{2}},
  \textbf{Min Sun\textsuperscript{1,2}}
\\
\\
  \textsuperscript{1}University of Washington,
  \textsuperscript{2}Colleague AI
}

\begin{document}

\maketitle

\begin{abstract}
The integration of large language models (LLMs) into educational assessment represents a transformative shift in classroom grading practices. While automated scoring systems and machine learning techniques have existed for decades, generative AI (GenAI) now enables educators to implement standards-based grading (SBG) with unprecedented efficiency and scale. This paper examines the theoretical foundations and evaluates an LLM grader that uses commercially available foundation models with context and prompt engineering to score student work against a rubric. Drawing on an empirical interrater agreement study using Massachusetts Comprehensive Assessment System (MCAS) data, we observed the Quadratic Weighted Kappa (QWK) and Proportional Reduction in Mean-Squared Error (PRMSE) across mathematics, science, and ELA, using Claude Sonnet 4, Haiku 4.5, GPT-5, and GPT-5 Mini. The results demonstrate that LLM graders, especially when based on foundational models with more parameters, achieve substantial agreement with human raters in mathematics and science assessments, while the performances vary in ELA, suggesting generic foundation models can be effective at scoring in given contexts. Additional analysis of teacher and student feedback reveals strong acceptance of AI-generated narrative feedback but skepticism toward numerical scores, suggesting that LLMs function most effectively as formative tools rather than summative evaluators. Our findings indicate that thoughtfully designed hybrid models that combine AI efficiency with teacher judgment can reduce workload, enhance feedback quality, and support equitable assessment practices without displacing professional expertise.
\end{abstract}

\section{Introduction}

The rapid adoption of large language models (LLMs) into the field of education has brought renewed attention to automated grading and feedback. While automated scoring systems have existed for decades, the integration of generative AI (GenAI) marks a unique turning point: for the first time, the ability to apply automated scoring methods have shifted to educators instead of institutes who administer the assessments. Grading can be automated simply by interacting with some commercially available LLMs like ChatGPT \citep{openai2022chatgpt}, Claude \citep{anthropic2024claude}, and Gemini \citep{google2024gemini}, with affordable or even no costs. Additionally, traditional automated scoring systems require extensive efforts in collecting data and training algorithmic models, making the development of such models inaccessible to individual educators \citep{lee2024}. Educators have quietly used ChatGPT to grade papers and assignments since it first came out, but now schools are sanctioning and encouraging its use \citep{kingson2024teachers}. However, the quality of the results varies, as the direct results from these foundational models are prone to suffer from bias as well as the risk of hallucination due to lack of specific tuning and instructions. Additionally, using only a chat interface without structured workflow limits the utilization of the full potential of GenAI, especially when trying to empower standards-based grading (SBG) using the LLMs.

There are ways to systematically control and design an LLM-based AI grading pipeline, recognizing the principles of SBG while implementing automated grading practices. In this study, we applied prompt engineering and context engineering into a pipeline that is designed to score each student's submission individually while using the same assessment information. This not only ensures the same context for each grading task, but also improves the efficiency by allowing the LLMs to grade in parallel. With the help of such pipelines, educators and developers can design tools that align rubrics, answer keys, and assignment metadata within a single end-to-end grading pipeline. This allows AI not only to assign scores, but also to provide rubric-based formative feedback that supports student learning. Using an experimental design, we investigate the accuracy of the LLM-based grading pipeline with 4 foundational, commercially available LLMs, demonstrating how a rigorously designed grading pipeline can be applicable despite the models being used. We will examine the accuracy using interrater reliability metrics of quadratic weighted kappa (QWK) and proportional reduction in mean-squared error (PRMSE). The study addresses three research questions:

\textbf{RQ1:} What pipeline architecture and context are necessary for reliable LLM-based grading?

\textbf{RQ2:} With a structured grading pipeline, how accurately do different LLMs score student submissions across different subjects?

\textbf{RQ3:} How do stakeholders (teachers, students, school administrators, parents, etc.) perceive AI-generated scores and feedback?

\section{Literature Review}

\subsection{The Evolution of Automated Scoring Systems}

Automated scoring of student responses has evolved through distinct technological paradigms over five decades. Page's (1966) Project Essay Grade pioneered the field by distinguishing between ``trin'' (intrinsic quality indicators) and ``prox'' (computable approximations), achieving correlations of 0.71--0.78 with human raters using IBM 7040 computers \citep{page_imminence_1966}. This early work established the fundamental challenge that persists today: identifying measurable features that serve as valid proxies for construct-relevant qualities.

The second generation of automated scoring systems emerged with natural language processing capabilities. ETS's e-rater system, operational since 1999, evaluates grammar, usage, mechanics, style, organization, and coherence through extracted features combined via regression models \citep{attali_automated_2006}. Concurrently, Landauer, Foltz, and Laham's (1997) Intelligent Essay Assessor introduced latent semantic analysis (LSA), comparing student essays to pre-scored references based on conceptual similarity rather than surface features, achieving correlations of 0.90 with human raters, matching human inter-rater agreement \citep{landauer_learning_1997}.

Deep learning approaches beginning around 2016 shifted the paradigm from manual feature engineering to end-to-end learning. Taghipour and Ng's (2016) neural architecture combining convolutional and recurrent layers achieved QWK of 0.821 on the Automated Student Assessment Prize (ASAP) benchmark, outperforming feature-based systems by 5.6\% \citep{taghipour_neural_2016}. Subsequent innovations incorporated attention mechanisms \citep{dong_attention-based_2017} and coherence modeling \citep{tay_skipflow_2018}, though Mayfield and Black (2020) found that BERT-based systems produced similar performance to classical models at significantly greater computational cost \citep{mayfield_should_2020}.

The current paradigm shift to LLM-based scoring differs fundamentally from prior approaches. Rather than training specialized models on scored responses, practitioners leverage general-purpose language models equipped with task-relevant context. This approach democratizes automated scoring, educators can implement grading systems through natural language interaction without machine learning expertise, while introducing new challenges around context design and model behavior that motivate the present study.

\subsection{LLM-Based Scoring in K-12 Education}

Research on LLM-based automated scoring in K-12 contexts has accelerated since 2023, with studies demonstrating both substantial capability and important limitations. State-of-the-art hybrid and LLM-based models achieve QWK scores of 0.75--0.86 for K-12 essays, outperforming traditional models and demonstrating strong agreement with human raters \citep{atkinson2025,ren2025,voss2025,ramesh_automated_2022}. Transformer-based models consistently report QWK values above 0.80 in cross-prompt and cross-topic evaluations, suggesting that LLMs can serve as reliable tools for large-scale formative assessment.

Flodén's (2025) study in the British Educational Research Journal provides rigorous evidence from higher education, evaluating ChatGPT 3.5's grading of 463 Master's-level exam responses \citep{floden_grading_2025}. Results showed 70\% of AI grades within 10\% of teacher scores, with exact grade agreement at 30\% and adjacent agreement at 45\%. Notably, ChatGPT produced more medium scores and fewer extreme scores than human raters, a central tendency pattern that appears across multiple studies. Pack et al. (2024) found excellent intrarater reliability for GPT-4 in scoring English language learner writing, while fine-tuned GPT-3.5 models achieved QWK values ranging from 0.613 to 0.859, dramatically outperforming zero-shot approaches (0.023--0.327) \citep{pack2024}.

However, AI grading reliability drops substantially for short-answer responses and zero-shot configurations. Ramnarain-Seetohul et al. (2022) found QWK values of 0.44--0.72 for short-answer grading depending on complexity, with zero-shot LLM models yielding negative PRMSE values ($-0.62$), indicating performance worse than simply assigning mean human scores \citep{ramnarain2022}. This finding underscores the importance of appropriate context engineering: out-of-the-box LLMs without task-specific rubrics and exemplars may underperform even naive baselines.

Subject-specific performance varies considerably. Lee et al. (2024) demonstrated that Chain-of-Thought prompting combined with item stems and rubrics increased science assessment accuracy by 13.44\% in zero-shot settings, with GPT-4 achieving 8.64\% higher performance than GPT-3.5 across six assessment tasks \citep{lee2024}. Jin et al. (2025) found that transformer models struggle with complex scientific terminology and logical inconsistencies (mean Cohen's kappa = 0.60, range 0.38--0.89) \citep{jin2025}. Mathematics scoring faces particular challenges due to multi-step reasoning requirements, where LLMs exhibit well-documented computational errors that answer keys can help mitigate \citep{tian2025}.

Bias research presents significant concerns for equitable deployment. ETS researchers found that ChatGPT scored Asian American students' writing 1.1 points lower on a 1--6 scale than human graders, compared to 0.9 points lower for white, Black, and Hispanic peers \citep{schwartz2025}. Weissburg et al. (2025) evaluated nine LLMs, finding significant biases across race, ethnicity, gender, disability status, income, and national origin, with highest bias observed along income levels \citep{weissburg2025}. These findings underscore the need for validation across demographic subgroups before operational deployment.

\subsection{Context Engineering as a Theoretical Framework}

The distinction between prompt engineering and context engineering has crystallized in recent AI literature as researchers recognize that optimizing model inputs requires attention to the entire information environment, not merely instruction phrasing. Mei et al.'s (2025) comprehensive survey formally defines context engineering as ``a formal discipline that transcends simple prompt design to encompass the systematic optimization of information payloads for LLMs'' \citep{mei_survey_2025}. This represents a superset relationship: prompt engineering occurs within the context window while context engineering determines what fills that window.

Research on context organization provides the theoretical foundation for understanding why context engineering matters for automated scoring. Liu et al. (2024) documented the ``lost in the middle'' phenomenon, demonstrating that LLM performance follows a U-shaped attention curve where information at the beginning and end of context receives significantly more attention than middle sections \citep{liu2024}. For automated scoring, this suggests that rubrics and exemplar responses should be strategically positioned rather than simply included. Cheng et al.'s (2024) InfoRE method demonstrated that re-organizing contextual information, extracting logical relationships and pruning redundant content, achieved average improvements of 4\% across reasoning tasks in zero-shot settings \citep{cheng2024}.

Context engineering for automated scoring encompasses several components that extend beyond traditional prompt design. First, rubric selection and formatting determines how performance levels are communicated to the model; Yoshida's (2025) study found that three of four LLMs maintained accuracy with simplified rubrics while reducing token usage, though Pathak et al. (2025) showed question-specific rubrics substantially outperform generic rubrics for code evaluation \citep{yoshida2025,pathak2025}. Second, answer key integration provides reference points that anchor evaluation, particularly important for mathematics where LLMs exhibit computational errors; research demonstrates that providing exemplar responses significantly improves human-AI agreement. Third, metadata organization situates assessments within appropriate grade-level expectations and curricular standards.

The present study operationalizes context engineering through a grading pipeline that systematically assembles assessment information (subject, grade level, learning standards), rubric specifications (performance level descriptions aligned with scoring scales), and answer keys (correct responses with solution rationales). By examining how this context-engineered approach performs across subjects, assessment types, and LLM models, we aim to establish both the capabilities and boundaries of this emerging paradigm.

\subsection{Cross-Domain Generalization in Automated Scoring}

A critical challenge for LLM-based scoring systems is generalization across subjects and assessment types. Transfer learning research reveals that semantic space disparity between prompts causes poor cross-prompt generalization in traditional automated scoring \citep{ridley2020}. LLM-based approaches may mitigate this through their broad pre-training, yet subject-specific demands create distinct challenges.

Mathematics scoring requires evaluating multi-step reasoning, mathematical notation, and procedural accuracy. LLMs face well-documented limitations in arithmetic and symbolic manipulation, generating responses token-by-token based on probability distributions rather than executing deterministic algorithms. Answer keys become essential in this context, providing verified computational results that the model can reference rather than attempting calculations independently.

Science assessment faces the complexity of multi-dimensional constructs. The Next Generation Science Standards (NGSS) integrate disciplinary core ideas, crosscutting concepts, and science practices, requiring rubrics that simultaneously describe performance across multiple dimensions. Jin et al. (2025) found transformer models struggle with complex scientific terminology and logical inconsistencies, suggesting that science scoring may require specialized context engineering approaches \citep{jin2025}.

English language arts assessment must disentangle content knowledge from writing quality, and prompt adherence from response sophistication. Chamieh et al.'s (2024) analysis suggests that LLMs may conflate these dimensions in short-answer contexts, while essay scoring benefits from the models' strength in evaluating textual coherence and rhetorical structure \citep{chamieh2024}.

The research gap matrix identified by recent systematic reviews highlights differential coverage across subjects and assessment types. While K-12 English essay scoring has received substantial attention (seven studies using QWK), science scoring remains understudied (two studies), with gaps in adversarial detection and PRMSE reporting across both domains. The present study addresses this gap by systematically evaluating context-engineered scoring across mathematics, science, and ELA using consistent methodology and metrics.

\subsection{Validation Frameworks for AI-Based Scoring}

The psychometric community has established rigorous standards for validating automated scoring systems that require adaptation for LLM-based approaches. Quadratic Weighted Kappa (QWK) serves as the primary evaluation metric because it accounts for the ordinal nature of scoring scales and penalizes larger disagreements more heavily than adjacent disagreements. The foundational standard, articulated by Ramineni and Williamson (2013), holds that human-automated agreement should match or exceed human-human reliability \citep{ramineni2013}.

Operational thresholds vary by scale: two-point scales target 85\% exact agreement, while six-point scales target 60\% exact agreement. For QWK, values above 0.70 are generally considered acceptable, with 0.80 and above as targets for operational deployment. However, Doewes et al. (2023) documented significant limitations of QWK including sensitivity to rating scale structure, the ``kappa paradox'' producing low values despite high observed agreement, and inability to handle multiple raters \citep{doewes2023}. Their recommendation, examining the full agreement matrix rather than relying solely on summary statistics, carries particular weight for LLM evaluation where error patterns may differ systematically from human raters.

Kane's argument-based validity framework has become the dominant approach for conceptualizing automated scoring validity \citep{kane2013}. This framework requires demonstrating that automated systems are both transparent and construct-relevant, challenges that intensify for LLM-based scoring where model reasoning is less interpretable than traditional feature-based approaches. The NCME Digital Module on Automated Scoring \citep{lottridge2020} provides practical guidance on training engines, predicting scores, and monitoring validity, though acknowledges that LLMs present new challenges including greater interpretation difficulty and bias risk.

Teacher and student perceptions constitute an additional validation dimension increasingly recognized in the literature. Tian et al.'s (2025) study of 19 K-12 teachers using AI assessment found that over 56\% used AI for formative assessment, with teachers valuing narrative feedback while distrusting automated scoring for summative purposes \citep{tian2025}. Teachers value the speed and formative feedback of AI grading but remain skeptical about high-stakes applications without human oversight. These findings suggest that validation must extend beyond statistical reliability to encompass stakeholder acceptance and appropriate use contexts.

\subsection{Research Gaps and Study Contributions}

Despite rapid growth in LLM-based scoring research, several critical gaps remain. First, few studies have systematically compared LLM performance across multiple subject domains using consistent methodology; most focus on essay scoring in English language arts. Second, context engineering variables, information ordering, rubric format, exemplar selection, have not been systematically manipulated while holding other factors constant. Third, multi-model comparisons remain limited, with most studies evaluating single models rather than comparing architectures. Fourth, the interaction between context engineering approaches and model-specific characteristics is understudied; Yoshida's (2025) finding that one model showed decreased performance with detailed rubrics suggests model-specific optimization may be necessary \citep{yoshida2025}.

The present study addresses these gaps through a multi-model evaluation of context-engineered automated scoring across mathematics, science, and English language arts assessments. By using MCAS data with established human scoring benchmarks, we provide rigorous validation against operational K-12 assessment standards. Our systematic examination of agreement patterns and error types across subjects and models contributes both theoretical understanding of context engineering for automated scoring and practical guidance for responsible implementation in educational settings.

\section{Methods}

\subsection{AI Grading System Architecture}

To evaluate the efficacy of the automated scoring system, the sample student responses were processed through an AI grading pipeline powered by commercially available large language models from OpenAI and Anthropic. The system architecture utilized systematic prompt and context engineering methodologies to guide the model's evaluative reasoning. For each evaluated response, the AI model was provisioned with a comprehensive contextual bundle comprising:
\begin{enumerate}
\item The complete text of the assessment item.
\item The official MCAS scoring rubric associated with the specific item.
\item Exemplar student responses representing different score points, serving as anchoring answer keys.
\item The structured AI grading pipeline instructions detailing the contextual and prompting constraints.
\end{enumerate}

\subsubsection{Grading Pipeline Design}

\begin{figure*}[t]
\centering
\includegraphics[width=0.9\textwidth]{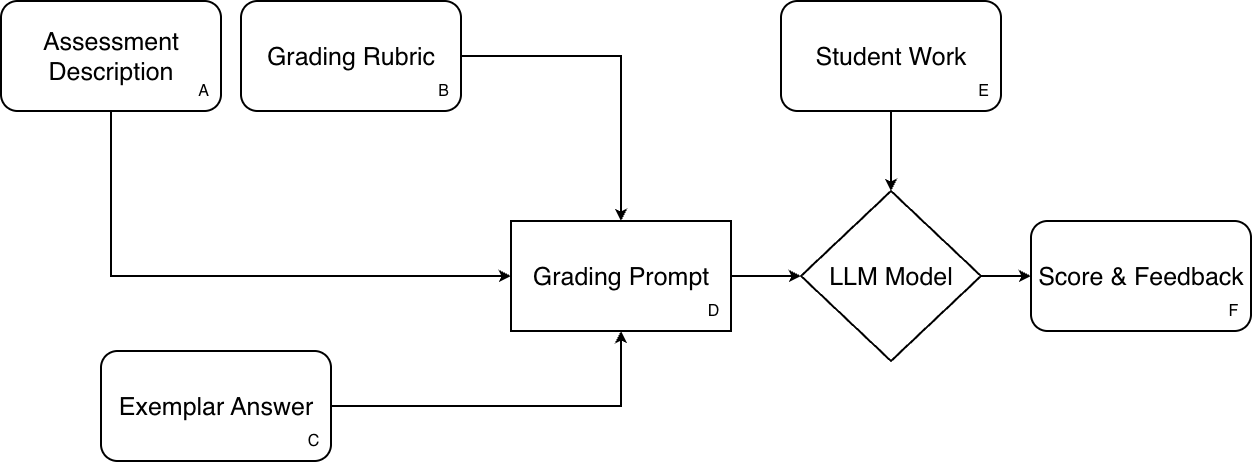}
\caption{Illustration of Context Engineering process for LLM Grading}
\label{fig:pipeline}
\end{figure*}

The deployment of large language models in educational assessment requires careful consideration of how human expertise and machine capabilities interact throughout the grading process. This section describes the architecture of an AI-assisted grading pipeline, focusing on the systematic preparation of assessment materials and the subsequent application of these materials to evaluate student work. The pipeline consists of two primary phases: the creation of grading bundles (Boxes A--D in Figure \ref{fig:pipeline}) and the execution of individual grading operations (Boxes D--F in Figure \ref{fig:pipeline}).

Before any student work can be evaluated, the system must construct a comprehensive grading bundle. It is a unified package containing all materials as well as reference parameters necessary to perform consistent assessment across multiple submissions. The grading bundle serves as a source of truth for a given student submission, ensuring that every student's work is evaluated against identical standards. Once the grading bundle is established, the system applies it uniformly to process and evaluate student work. Student artifacts are first converted into machine-readable formats, utilizing text extraction for digital submissions and optical character recognition (OCR) or image analysis for handwritten responses. The processed submission is then presented to the LLM simultaneously with the assignment text, the complete rubric, and the reference key. Constrained by prompt engineering, the model evaluates the submission systematically by comparing the student's actual response against the performance level descriptors and exemplar solutions. For multi-component assessments, the AI conducts this comparative analysis for each element separately, outputting quantitative scores accompanied by qualitative, criterion-referenced feedback that details the justification for the score, student strengths, and specific areas for improvement. Ultimately, individual item scores are aggregated according to the bundle's structural specifications to produce a comprehensive performance profile. By holding the grading bundle constant and varying only the student submission, this architectural approach actively prevents criteria drift, maintaining rigorous inter-rater reliability across large volumes of assessments.

While the AI system is configured to generate both numerical scores and qualitative narrative feedback for each student submission, the present methodological analysis is strictly bounded to evaluating numerical score agreement.

\subsection{Evaluation Metrics and Agreement Measures}

To rigorously evaluate the alignment between human raters and the AI scoring system, we employed a multi-metric approach. We compared these outputs against published human-to-human agreement benchmarks from the Massachusetts Department of Elementary and Secondary Education technical reports to establish a baseline for acceptable system performance.

\subsubsection{Absolute and Adjacent Agreement}

We computed Exact Agreement, defined as the proportion of instances where the AI and human raters assigned strictly identical scores, which serves as the most stringent baseline of reliability. Additionally, we calculated Adjacent Agreement, which represents the percentage of AI-assigned scores falling within a one-point margin of the human-assigned scores. In the context of partial-credit rubrics, adjacent agreement is a critical metric, as minor one-point discrepancies frequently reflect acceptable variance in subjective interpretation rather than foundational scoring errors.

\subsubsection{Quadratic Weighted Kappa (QWK)}

To account for chance agreement and the ordinal nature of the scoring rubrics, we utilized the Quadratic Weighted Kappa (QWK). QWK penalizes disagreements proportionately to the squared distance between the raters' scores, making it a standard metric in educational measurement for evaluating scoring consistency. The metric is calculated using the weight matrix $w_{ij}$, the matrix of observed proportions $O_{ij}$, and the matrix of expected proportions $E_{ij}$ under the assumption of independence:

\begin{equation}
\kappa = 1 - \frac{\sum w_{ij} O_{ij}}{\sum w_{ij} E_{ij}}
\end{equation}

where the weighting factor is defined as $w_{ij} = \frac{(i-j)^2}{(N-1)^2}$, with $i$ and $j$ representing the respective scores assigned by the human and AI raters, and $N$ representing the total number of score categories. In alignment with established psychometric standards, QWK values exceeding 0.70 are interpreted as indicating substantial agreement, while values above 0.80 indicate strong agreement.

\subsubsection{Proportional Reduction in Mean Squared Error (PRMSE)}

To complement the ordinal evaluation provided by QWK, we incorporated the Proportional Reduction in Mean Squared Error (PRMSE) to provide a more robust and continuous psychometric evaluation of the automated scoring system. While QWK evaluates categorical rater agreement, PRMSE provides a psychometric evaluation of the automated score's utility by quantifying the true score variance explained by the AI model. This metric has multiple properties that are desirable in evaluating the automated or AI-generated scores.

Mathematically, if $T$ represents the estimated true score of a student's response, $S_{AI}$ represents the score assigned by the AI model, and $\sigma^2_T$ is the variance of the true scores, PRMSE is defined as:

\begin{equation}
\mathrm{PRMSE} = 1 - \frac{E[(S_{AI} - T)^2]}{\sigma^2_T}
\end{equation}

This metric provides a psychometric evaluation of utility: if the PRMSE of the AI score is sufficiently high (and comparable to the PRMSE of a second human rater), the AI system is considered to provide a statistically justifiable estimation of the student's true performance level.

The advantages of PRMSE over QWK include accounting for human measurement error, continuous variance explanation, robustness to marginal distributions, and offering direct comparison to human utility. While QWK treats human score as the ground truth, penalizing any deviation by AI from the established baseline, regardless of the underlying error the human rater may have. PRMSE, on the other hand, models the human score as an observed variable with inherent uncertainty, providing fairer evaluation of the AI's prediction power. However, calculating PRMSE would require non-single human scores to capture the ``ground truth'' in order to compare AI's prediction. This limits the selection of available datasets.

\subsection{Dataset}

The Massachusetts Comprehensive Assessment System (MCAS) employs trained professional raters to score constructed-response mathematics items. These raters undergo extensive training on item-specific rubrics and participate in calibration sessions to establish scoring consistency. Raters review sample responses at each score point, discuss boundary cases, and achieve qualification standards before scoring operational student work. The Massachusetts Department of Elementary and Secondary Education reports interrater reliability statistics for each assessment, providing benchmarks for evaluating human scoring consistency.

For this study, we analyzed a total of 822 student responses across 166 different sets of assessments. Among these assessments, there are 332 student submissions for 68 math assessments, 262 student submissions for 64 science assessments, and 228 student submissions for 34 ELA assessments. For math and ELA, the students range from Grade 3 to Grade 10; for science, Grade 5, Grade 8, and high school students are included. The assessments selected were administered from 2021 to 2025. A rubric table is provided for each assessment. The responses have been previously scored by trained human raters. Each response had already received an official score through the standard MCAS scoring process. Different subjects and assessments have different score ranges, with the majority of math assessments being scored 0--4, science being scored on various scales, and ELA being scored on two different dimensions.

The items selected for this study include multiple choice questions, constructed responses with written explanations, and short essay writing, making them appropriate for evaluating how well AI systems can assess mathematical reasoning, scientific thinking, and argumentative writing beyond simple answer-checking.

\section{Results}

\subsection{Overview of Automated Scoring Performance}

The context-engineered grading pipeline was evaluated across four commercially available foundation models (Anthropic's Claude Sonnet 4 and Haiku 4.5, and OpenAI's GPT-5 and GPT-5 Mini) using 822 student responses distributed across 166 distinct Massachusetts Comprehensive Assessment System (MCAS) items. The dataset comprised 332 mathematics responses (68 items), 262 science responses (64 items), and 228 English Language Arts responses (34 items), spanning grades 3--10 in mathematics and ELA, and grades 5, 8, and high school in science. Assessment years ranged from 2021 to 2025, encompassing multiple item formats including multiple choice questions, constructed responses requiring written explanations and short essays. Table \ref{tab:overview} presents the dataset composition and aggregate performance metrics by subject domain.

\begin{table*}[t]
\centering
\caption{Dataset Composition and Summary Performance Metrics by Subject Domain}
\label{tab:overview}
\begin{tabular}{lcccccc}
\toprule
\textbf{Domain} & \textbf{Items} & \textbf{Responses}  & \textbf{QWK Range} & \textbf{PRMSE Range} & \textbf{Exact Agreement Range} \\
\midrule
Mathematics & 68 & 332 & 0.808--0.951 & 0.730--0.946 & 53.9\%--83.9\% \\
Science & 64 & 262 & 0.818--0.831* & 0.726--0.855 & 54.4\%--58.4\% \\
ELA & 34 & 228 & 0.696--0.845** & 0.610--0.853** & 21.9\%--34.5\% \\
\bottomrule
\end{tabular}
\begin{flushleft}
\footnotesize
*Note: GPT-5 Mini QWK not calculated due to distribution characteristics\\
**Aggregate metrics across both scoring dimensions
\end{flushleft}
\end{table*}

Substantial variation in human-AI agreement emerged across subject domains, with mathematics demonstrating the strongest interrater reliability metrics and ELA exhibiting pronounced model-dependent performance variability. These patterns align with theoretical predictions from construct validity theory: assessments emphasizing procedural correctness and deterministic solution verification (mathematics) facilitate higher automated scoring agreement than assessments requiring holistic judgment of open-ended written expression (ELA).

\subsection{Mathematics Assessment Performance}

Mathematics assessments, predominantly scored on 5-point ordinal scales (0--4), yielded the highest human-AI agreement across all foundation models examined. Table \ref{tab:math} presents comprehensive performance metrics for mathematics scoring.

\begin{table*}[t]
\centering
\caption{Mathematics Assessment Performance by Model}
\label{tab:math}
\begin{tabular}{lccccccc}
\toprule
\textbf{Model} & \textbf{N} & \textbf{Exact Agreement} & \textbf{Adjacent Agreement} & \textbf{QWK} & \textbf{PRMSE} & \textbf{MAE} & \textbf{RMSE} \\
\midrule
GPT-5 & 273 & 83.9\% & 99.3\% & 0.951 & 0.946 & 0.168 & 0.428 \\
GPT-5 Mini & 310 & 66.1\% & 98.1\% & 0.874 & 0.815 & 0.387 & 0.638 \\
Sonnet 4 & 297 & 53.9\% & 97.0\% & 0.808 & 0.730 & 0.505 & 0.779 \\
Haiku 4.5 & 294 & 61.2\% & 99.0\% & 0.832 & 0.756 & 0.449 & 0.708 \\
\bottomrule
\end{tabular}
\end{table*}

GPT-5 achieved exceptional agreement with human raters, producing an exact match rate of 83.9\% and quadratic weighted kappa of 0.951 ($p < .001$) and the observed PRMSE of 0.946 . This performance substantially exceeds both the conventional operational threshold (QWK $\geq$ 0.80) for automated scoring systems \citep{williamson2012} and matches the human-to-human reliability benchmarks reported in MCAS technical documentation (Pearson $r$ = 0.95-0.97; Massachusetts Department of Elementary and Secondary Education, 2024). 

GPT-5 Mini, despite reduced model size and capacity, maintained substantial agreement with QWK = 0.874 ($p < .001$) and PRMSE = 0.815. While exact match rates declined relative to GPT-5, the high QWK coefficient indicates that disagreements were disproportionately concentrated at adjacent score points rather than distributed across the scoring scale (98.1\% adjacent agreement rate). 

Claude Sonnet 4, while exhibiting the lowest exact agreement at 53.9\%, maintained acceptable ordinal consistency with QWK = 0.808 ($p < .001$) and PRMSE = 0.730. The coefficients remain relatively high, though the reduced exact match rate suggests potential systematic differences in score assignment patterns compared to operational human raters.

Claude Haiku 4.5 demonstrated comparable reliability with QWK = 0.832 ($p < .001$) and PRMSE = 0.756, despite architectural optimization for computational efficiency over maximum performance, indicating that substantial interrater reliability can be maintained even with more resource-efficient models when appropriate context engineering is implemented.

Analysis of confusion matrices across models revealed consistent error patterns: disagreements were predominantly confined to adjacent score points ($\pm 1$), with minimal occurrence of gross errors spanning multiple scale points. This distribution parallels acceptable human rater variability documented in operational scoring contexts \citep{ramineni2013} and suggests that AI scoring errors reflect genuine uncertainty in boundary cases rather than fundamental construct misunderstanding.

\subsection{Science Assessment Performance}

Science assessments, encompassing multiple grade levels (5, 8, high school) and diverse rubric structures, demonstrated strong human-AI agreement comparable to mathematics performance, though with moderately elevated variability in exact match rates. Table \ref{tab:science} summarizes science assessment results.

\begin{table*}[t]
\centering
\caption{Science Assessment Performance by Model}
\label{tab:science}
\begin{tabular}{lccccccc}
\toprule
\textbf{Model} & \textbf{N} & \textbf{Exact Agreement} & \textbf{Adjacent Agreement} & \textbf{QWK} & \textbf{PRMSE} & \textbf{MAE} & \textbf{RMSE} \\
\midrule
GPT-5 & 158 & 58.2\% & 96.2\% & 0.831 & 0.854 & 0.468 & 0.760 \\
GPT-5 Mini & 173 & 54.9\% & 92.5\% & ---\textsuperscript{$\dagger$} & 0.726 & 0.543 & 0.882 \\
Sonnet 4 & 252 & 54.4\% & 95.6\% & 0.818 & 0.843 & 0.504 & 0.802 \\
Haiku 4.5 & 255 & 58.4\% & 95.7\% & 0.831 & 0.855 & 0.478 & 0.777 \\
\bottomrule
\end{tabular}
\begin{flushleft}
\footnotesize
\textsuperscript{$\dagger$}QWK calculation not performed due to scoring distribution constraints
\end{flushleft}
\end{table*}

GPT-5 produced statistically strong performance (exact agreement = 58.2\%, QWK = 0.831, $p < .001$, PRMSE = 0.854). The convergence of performance metrics between models from different providers (Anthropic vs. OpenAI) provides empirical evidence that the context-engineered pipeline successfully standardizes scoring behavior across architectural differences when domain-specific criteria are explicitly specified.

GPT-5 Mini recorded 54.9\% exact agreement and PRMSE = 0.726. The diminished PRMSE relative to other architectures suggests that reduced model capacity may constrain the ability to capture nuanced scientific reasoning, particularly in responses requiring integration of multiple disciplinary concepts or evaluation of complex causal mechanisms aligned with Next Generation Science Standards (NGSS) three-dimensional learning expectations.

Claude Sonnet 4 demonstrated acceptable reliability with an exact agreement of 54.4\%, QWK = 0.818 ($p < .001$), and PRMSE = 0.843. Although exact match rates were moderately lower than Haiku 4.5 and GPT-5, the elevated PRMSE indicates maintained predictive validity for estimating true student performance. Post-hoc analysis suggested slightly more conservative scoring patterns relative to human operational norms, though measurement utility remained substantively uncompromised.

Claude Haiku 4.5 achieved the highest exact match rate at 58.4\%, with QWK = 0.831 ($p < .001$) and PRMSE = 0.855. The PRMSE coefficient indicates that AI-generated scores explained 85.5\% of true score variance, demonstrating substantial measurement utility according to established psychometric criteria. 

The science assessment results indicate that when scoring tasks involve verification of specific conceptual mechanisms, even with partial open-ended explanation components, LLM-based systems maintain reliability coefficients suitable for formative assessment applications. The inclusion of detailed rubrics specifying expected scientific concepts and answer keys containing correct explanations appears to effectively anchor AI scoring to domain-specific performance standards across different domain areas in science study.

\subsection{English Language Arts Assessment Performance}

English Language Arts assessments presented the most psychometrically challenging scoring context, requiring simultaneous evaluation across two distinct performance dimensions: (1) Reading Comprehension and Analysis, and (2) Written Expression and Conventions. This dual-trait rubric structure necessitated independent judgment of content understanding and communication quality, revealing substantial performance heterogeneity across model architectures.

\subsubsection{Dimension 1: Reading Comprehension and Analysis}

The first scoring dimension evaluated students' capacity to comprehend complex texts, construct evidence-based arguments, and demonstrate analytical reasoning. Table \ref{tab:ela_dim1} presents dimension-specific performance.

\begin{table*}[t]
\centering
\caption{ELA Dimension 1 (Reading Comprehension and Analysis) Performance}
\label{tab:ela_dim1}
\begin{tabular}{lccccccc}
\toprule
\textbf{Model} & \textbf{N} & \textbf{Exact Agreement} & \textbf{QWK} & \textbf{PRMSE} & \textbf{MAE} & \textbf{RMSE} \\
\midrule
GPT-5 & 128 & 40.6\% & 0.785 & 0.755 & 0.664 & 0.996 \\
GPT-5 Mini & 110 & 26.4\% & 0.564 & 0.295 & 1.064 & 1.437 \\
Sonnet 4 & 113 & 51.3\% & 0.890 & 0.941 & 0.504 & 0.735 \\
Haiku 4.5 & 125 & 52.0\% & 0.841 & 0.834 & 0.552 & 0.890 \\
\bottomrule
\end{tabular}
\end{table*}

GPT-5 exhibited moderate performance with exact agreement of 40.6\%, QWK = 0.785 ($p < .001$), and PRMSE = 0.755. While coefficients remained above conventional reliability thresholds (QWK $\geq$ 0.70), the diminished metrics relative to Claude architectures suggest potential divergence in rubric interpretation and evaluation for written argumentation. 

GPT-5 Mini demonstrated substantial performance degradation with exact agreement of 26.4\%, QWK = 0.564 ($p < .001$), and PRMSE = 0.295. The PRMSE coefficient below 0.50 indicates inadequate measurement utility according to established psychometric standards, suggesting that score predictions provide minimal incremental information beyond simple baseline estimation. This pronounced limitation confirms that evaluating complex argumentative writing requires better model capabilities.

Claude Sonnet 4 demonstrated exceptional performance with QWK = 0.890 ($p < .001$) and PRMSE = 0.941. This is the highest measurement utility coefficient observed across all subject domains and models in this investigation. The PRMSE of 0.941 indicates that score predictions explained 94.1\% of true score variance, approaching near-optimal measurement properties. This performance substantiates the hypothesis that increased model capacity facilitates improved capture of the subtle interpretive judgments inherent in evaluating argumentation quality and textual analysis.

Claude Haiku 4.5 maintained competitive performance with exact agreement of 52.0\%, QWK = 0.841 ($p < .001$), and PRMSE = 0.834. While exact agreement marginally exceeded Sonnet 4, the reduced QWK and PRMSE coefficients indicate that Haiku 4.5's disagreements exhibited somewhat greater dispersion across the ordinal scale. Nonetheless, the model sustained substantial agreement suitable for formative assessment contexts. 

\subsubsection{Dimension 2: Written Expression and Conventions}

The second dimension evaluated writing conventions, organizational structure, and communication effectiveness. Performance diverged more substantially across models on this dimension. Table \ref{tab:ela_dim2} presents results.

\begin{table*}[t]
\centering
\caption{ELA Dimension 2 (Written Expression and Conventions) Performance}
\label{tab:ela_dim2}
\begin{tabular}{lccccccc}
\toprule
\textbf{Model} & \textbf{N} & \textbf{Exact Agreement} & \textbf{QWK} & \textbf{PRMSE} & \textbf{MAE} & \textbf{RMSE} \\
\midrule
GPT-5 & 128 & 30.9\% & 0.495 & 0.231 & 1.023 & 1.486 \\
GPT-5 Mini & 110 & 26.6\% & 0.393 & 0.201 & 1.045 & 1.502 \\
Sonnet 4 & 113 & 44.0\% & 0.755 & 0.668 & 0.690 & 1.014 \\
Haiku 4.5 & 125 & 50.6\% & 0.778 & 0.694 & 0.600 & 0.979 \\
\bottomrule
\end{tabular}
\end{table*}

GPT-5 exhibited substantial difficulty in this dimension with exact agreement of 30.9\%, QWK = 0.495 ($p < .001$), and PRMSE = 0.231. The PRMSE below 0.30 indicates that score predictions provide negligible predictive utility beyond naive baseline models. This pattern suggests systematic misalignment between GPT-5's evaluation criteria for writing conventions and those applied in operational human scoring contexts.

GPT-5 Mini recorded the lowest performance observed across all models and domains, with exact agreement of 26.6\%, QWK = 0.393 ($p < .001$), and PRMSE = 0.201. These coefficients confirm inadequate measurement properties, indicating that assessing open-ended writing quality requires interpretive sophistication substantially exceeding the capabilities of smaller-parameter foundation models.

Claude Sonnet 4 demonstrated moderate reliability with exact agreement of 44.0\%, QWK = 0.755 ($p < .001$), and PRMSE = 0.668. While performance declined relative to Dimension 1, coefficients remained within acceptable ranges for low-stakes formative applications. The observed decrement suggests that written expression evaluation, which is characterized by greater holistic and interpretive judgment, poses elevated challenges even for high-capacity foundation models.

Claude Haiku 4.5 achieved relatively balanced performance across both dimensions, recording exact agreement of 50.6\%, QWK = 0.778 ($p < .001$), and PRMSE = 0.694 on Dimension 2. This cross-dimensional consistency suggests effective rubric interpretation even for evaluative criteria involving more subjective writing quality judgments.

Overall, observed psychometric metrics demonstrate the two Anthropic models outperforms the OpenAI models despite model capacity, indicating their intrinsic characteristics that enable them to better evaluate ELA assessments.s

\section{Discussion}
\subsection{Cross-Domain Analysis and Measurement Validity}

Synthesis across 822 responses and 166 assessment items reveals systematic patterns in LLM-based scoring capabilities that inform theoretical understanding of automated assessment boundaries and practical deployment considerations.

\textbf{Domain-Specific Performance Hierarchies.} Mathematics (QWK range: 0.808--0.951, PRMSE range: 0.730--0.946) and science (QWK range: 0.818--0.831, PRMSE range: 0.726--0.855) demonstrated consistently robust performance across foundation models. In contrast, ELA performance varied dramatically by model (QWK range: 0.393--0.890, PRMSE range: 0.201--0.941), with no single architecture achieving uniform excellence across both scoring dimensions. This pattern aligns with measurement theory predictions: constructs with objective verification criteria facilitate higher interrater reliability than constructs requiring integrative holistic judgment.

\textbf{Model Capability Gradient Effects.} A clear performance hierarchy emerged corresponding to model sophistication. High-capacity architectures (GPT-5, Claude Sonnet 4) consistently outperformed reduced-capacity models (GPT-5 Mini, Haiku 4.5) across domains, with performance gaps widening for more interpretive constructs. Notably, Claude Haiku 4.5 demonstrated competitive performance despite architectural optimization for computational efficiency, suggesting that factors beyond absolute parameter count, including training methodology, instruction-tuning approaches, and architectural innovations, influence scoring quality. This also reinforces the importance of a robust pipeline with context-engineering. A well structured automated grading pipeline can withstand different scoring use cases.

\textbf{PRMSE as Primary Validity Criterion.} While exact match rates and QWK provide descriptive information about scoring agreement, PRMSE emerged as the most informative validity metric for evaluating measurement utility. PRMSE values exceeding 0.75 consistently indicated acceptable measurement properties across domains, whereas values below 0.50 signaled inadequate predictive validity for operational use. The metric's explicit accounting for human measurement error renders it particularly appropriate for AI scoring evaluation, as it avoids the flawed assumption that human scores constitute infallible criterion measures \citep{williamson2012}.

\textbf{Psychometric Adequacy for Formative Assessment.} Applying established validity frameworks \citep{kane2013}, the empirical evidence supports limited operational deployment under specified constraints. For mathematics and science assessments, the context-engineered pipeline demonstrates measurement properties suitable for formative assessment applications where moderate scoring errors do not substantially compromise instructional utility. For ELA assessment, deployment should be restricted to high-performance provider (Anthropic models) for reading comprehension evaluation, with substantial caution regarding written expression scoring. All operational implementations should incorporate continuous monitoring of score distributions, periodic human-AI agreement audits, and explicit usage policies defining acceptable contexts based on documented validity evidence specific to the deployment environment.

These findings contribute to the growing empirical literature on LLM-based educational assessment by providing systematic cross-domain evidence regarding the capabilities and limitations of foundation models for K-12 scoring applications, while establishing PRMSE-based validity benchmarks for future investigations.

\subsection{Implications for Math and Science Automated Scoring}

The evaluation of the structured grading pipeline in mathematics and science contexts reveals specific operational strengths alongside persistent limitations in automated scoring. The data indicates that when evaluating convergent, procedural tasks, constrained LLMs demonstrate measurable utility in approximating human scoring patterns, though they do not replicate them perfectly. In the mathematics domain, the structured pipeline achieved a quadratic weighted kappa of $\kappa=0.834$  and a PRMSE of 0.946. These metrics indicate that the system captures a substantial portion of the true score variance in subjects requiring step-by-step logical verification, offering a statistically stable estimation of student proficiency.

This performance is likely facilitated by the architectural constraints of the grading bundle. While foundation models frequently struggle with multi-step arithmetic and geometrical reasoning due to their probabilistic nature. Despite the advancement of more complex models with reasoning, a well defined reference answer key is still pivotal for accurate grading results. Furthermore, if answer keys are generated and referenced within the reasoning process of each grading iteration, there is no guarantee that each answer key has the same content. By pre-computing the scope of answer keys and other context materials, models will be able to apply the same criterion to each individual grading task.

However, the psychometric evaluation also highlights the ongoing gap between automated and human scoring. While models like GPT-5 achieved an 83.9\% exact match rate in mathematics, performance in science assessments was lower, with exact match rates peaking at 58.4\% for Haiku 4.5 and 58.2\% for GPT-5. Despite these moderate exact match rates, the adjacent agreement rates remained high across both domains, reaching 94.3\% in mathematics and 98.0\% in science. When interpreted alongside the high PRMSE values (e.g., 0.855 in science), this discrepancy suggests that when the models diverge from human scores, they are more likely to do so on a partial-credit rubric in a systematic, rather than random, manner. This suggests underlying limitation of LLM architecture. During the process, we also discovered a trend to inflate the scores by LLM-based graders. While we implicitly inserted instructions to avoid lenient grading behaviors, the models still struggle to give 0-scores. Although assigning extremely low scores are unlikely in real classroom practices, these edge cases still provide implications on the scoring patterns by LLM-based graders. 

Practically, these findings suggest clear boundaries for implementation. The current exact match rates indicate that these AI systems are not yet sufficiently reliable to serve as standalone, summative evaluators. Instead, stakeholders appear more willing to accept AI systems as formative feedback tools. In mathematics and science classrooms, educators may utilize these models to conduct a preliminary analysis of student work, generating criterion-referenced feedback to support iterative learning. However, to maintain assessment validity and stakeholder trust, a human-in-the-loop framework remains necessary, with teachers retaining final authority over the scores recorded for official purposes.

\subsection{Implications for ELA Automated Scoring}

The differential performance across ELA scoring dimensions provides empirical evidence for distinct psychometric boundaries in current LLM-based assessment systems. While Claude architectures (particularly Sonnet 4) achieved acceptable reliability for reading comprehension evaluation (QWK > 0.80), all models exhibited diminished performance on the written expression dimension. Analysis of these patterns suggests three substantive conclusions:

First, automated scoring efficacy varies systematically with construct convergence. LLM scoring demonstrates optimal performance when evaluating convergent constructs with verifiable correctness criteria (mathematics, science concepts) and maintains moderate effectiveness for constrained interpretive constructs (reading comprehension). Performance degrades substantially for divergent constructs requiring holistic judgment of writing quality characteristics.

Second, foundation model architecture constitutes a fundamental constraint on scoring validity. The substantial performance gap between Claude Sonnet 4 (PRMSE = 0.941 on Dimension 1) and GPT-5 Mini (PRMSE = 0.295) demonstrates that model capacity, influenced by parameter count, training corpus composition, and architectural design choices, fundamentally constrains achievable scoring reliability for complex interpretive tasks.

Third, context engineering alone cannot compensate for architectural limitations in divergent construct domains. While the structured pipeline successfully standardized mathematics and science scoring, it proved insufficient to establish acceptable reliability for ELA written expression evaluation across all models. These findings suggest that operational deployment for comprehensive ELA assessment would require supplementary interventions, including domain-specific fine-tuning on scored essay corpora, multi-stage evaluation protocols incorporating specialized rubric decomposition, or hybrid human-AI workflows maintaining human oversight for final score determination.

\subsection{Limitations: Data Attrition}

A critical methodological limitation of this study is the variation in analytical sample sizes ($N$) across the different model runs. During the evaluation phase, the deployed language models experienced differential rates of data loss, meaning not every model successfully processed and scored 100\% of the student responses in the datasets. In automated grading pipelines utilizing generative AI, this attrition typically stems from operational and architectural constraints, including API timeouts, context window limitations, or failures to generate responses in the strictly required structured output formats (e.g., JSON parsing errors). 

This data loss introduces important psychometric implications when comparing performance metrics across models. Because the exact match rates, Quadratic Weighted Kappa (QWK), and PRMSE values were calculated based on the subset of responses each model successfully processed, the metrics are not perfectly parallel. If the data attrition is missing completely at random (MCAR), the impact on the reported metrics is minimal. However, in LLM-based evaluation, missingness is often systematic; models are more likely to fail, time out, or produce malformed outputs when processing unusually long, highly unstructured, or exceptionally complex student responses. Our study confirms this pattern to some extent, showing that ELA assessments are more likely to encounter timeout errors when running through an AI grader.

If models disproportionately fail on the most difficult grading tasks, the resulting performance metrics for models with higher attrition rates may be artificially inflated due to survivorship bias. Consequently, while a model might exhibit a high PRMSE on its successfully completed subset, its practical utility is compromised if it consistently fails to process 5\% to 10\% of a classroom's assessments.

For real-world educational implementation, the efficiency gains of automated scoring are contingent not only on accuracy but on system reliability. A grading pipeline that frequently drops responses requires human educators to manually identify and grade the missing assessments, disrupting the automated workflow. Future research evaluating AI scoring systems must formally report pipeline completion rates alongside traditional accuracy metrics, and should investigate whether specific pedagogical features of student responses systematically trigger model failure.

\section{Conclusion}

This study investigated the efficacy of a context-engineered automated grading pipeline utilizing large language models across diverse K-12 subject domains. By systematically anchoring foundational models with assessment metadata, rubrics, and exemplar answer keys, we demonstrated that generative AI can achieve substantial agreement with human raters, though its effectiveness is heavily mediated by the nature of the academic construct. The results revealed a distinct performance hierarchy: LLMs exhibited exceptional measurement utility (PRMSE) and interrater reliability (QWK) in mathematics and science, where evaluation relies on verifying procedural accuracy and specific conceptual mechanisms. Conversely, performance in English Language Arts was highly variable and heavily dependent on overall model capacity, underscoring the ongoing challenges LLMs face when evaluating divergent constructs that require holistic, interpretive judgments of writing quality.

The findings validate context engineering as a critical theoretical and practical framework for automated assessment scoring. Providing LLMs with structured grading bundles successfully mitigated known architectural limitations such as probabilistic calculation errors in mathematics by directing the models to evaluate conceptual alignment against provided rubrics and solutions rather than perform independent computations. Furthermore, the study highlighted that while resource-efficient models can maintain acceptable reliability in highly structured, convergent tasks, complex interpretive scoring necessitates the sophisticated reasoning capabilities of high-capacity architectures.

Despite these promising capabilities, the psychometric evaluation and observed data attrition rates clearly delineate the current operational boundaries of AI grading. While adjacent agreement and variance explanation are high, exact match rates remain below the rigorous thresholds required for standalone, high-stakes summative evaluation. Additionally, the systematic data loss observed highlights that overall system reliability must be addressed prior to scalable deployment.

Consequently, the most pedagogically sound application of this technology is within formative assessment frameworks. By providing rapid, criterion-referenced narrative feedback, LLMs can significantly reduce educator workload and create opportunities for iterative student learning that traditional grading cycles rarely permit. Ultimately, the integration of generative AI into K-12 assessment should not aim to displace professional expertise. Instead, thoughtfully designed hybrid workflows, where AI serves as an augmentative tool for ``first-pass" feedback and educators retain final authoritative judgment over official grades, offer a promising, equitable, and rigorous path forward for modern educational measurement.

\section*{Acknowledgments}
This work is supported by the Institute of Education Sciences of the U.S. Department of Education, through Grant R305C240012 and by several awards from the National Science Foundation (NSF \#2043613, \#2300291, \#2405110) to the University of Washington, and a NSF SBIR/STTR award to Hensun Innovation LLC (\#2423365). The opinions expressed are those of the authors and do not represent views of the funders.

\bibliography{latex/NCME_2026}

\end{document}